\documentclass[preprint, 12pt]{elsarticle}
\usepackage{graphicx}
\usepackage{amssymb}
\usepackage[fleqn]{amsmath}
\usepackage{lineno}
\usepackage{color}

\journal{arXiv}

\usepackage[section]{placeins}

\usepackage{framed}
\usepackage{multicol}
\usepackage{multirow}
\usepackage{makecell}
\usepackage{nomencl}
\usepackage{amsmath}
\usepackage[english]{babel}
\usepackage{hyphenat}
\usepackage{algorithm,algcompatible}
\usepackage{varwidth}

\begin{document}
\begin{frontmatter}


%\title{Assessing the impact of topographic uncertainty on flow-like landslide run-out modeling using unconditional and conditional stochastic simulation}

% just playing with the title ... 
\title{Emulator-based global sensitivity analysis for flow-like landslide run-out models}



\author[AICES]{Hu Zhao}
%\ead{zhao@aices.rwth-aachen.de}

\author[LIH]{Florian Amann}

\author[AICES,Göttingen]{Julia Kowalski}
%\ead{kowalski@aices.rwth-aachen.de}

\address[AICES]{AICES Graduate School, RWTH Aachen University, Schinkelstr. 2a, 52062 Aachen, Germany}
\address[LIH]{Department of Engineering Geology and Hydrogeology, RWTH Aachen University, Lochnerstr. 4-20, 52056 Aachen, Germany}
\address[Göttingen]{Computational Geoscience, Geoscience Centre, University of Göttingen, Goldschmidtstr. 1, 37077 Göttingen, Germany}

\begin{abstract}
Landslide run-out modeling involves various uncertainties originating from model input data. It is therefore desirable to assess the model's sensitivity. A global sensitivity analysis that is capable of exploring the entire input space and accounts for all interactions, often remains limited due to computational challenges resulting from a large number of necessary model runs. We address this research gap by integrating Gaussian process emulation into landslide run-out modeling and apply it to the open-source simulation tool r.avaflow. The feasibility and efficiency of our approach is illustrated based on the 2017 Bondo landslide event. The sensitivity of aggregated model outputs, such as the apparent friction angle, impact area, as well as spatially resolved maximum flow height and velocity, to the dry-Coulomb friction coefficient, turbulent friction coefficient and the release volume are studied. The results of first-order effects are consistent with previous results of common one-at-a-time sensitivity analyses. In addition to that, our approach allows to rigorously investigate interactions. Strong interactions are detected on the margins of the flow path where the expectation and variation of maximum flow height and velocity are small. The interactions generally become weak with increasing variation of maximum flow height and velocity. Besides, there are stronger interactions between the two friction coefficients than between the release volume and each friction coefficient. In the future, it is promising to extend the approach for other computationally expensive tasks like uncertainty quantification, model calibration, and smart early warning.
\end{abstract}


\end{frontmatter}

%\linenumbers

%% 1. Introduction
\section{Introduction}
\label{S:1}
Flow-like landslides, e.g. rock avalanches and debris flows, pose an ongoing threat to life, property, and environment in mountainous regions around the world. In order to assess their hazard and design mitigation strategies, many research efforts have been devoted to developing computational landslide run-out models which are capable of simulating the dynamics of the flow over complex topographies. The majority of these models employ depth-averaged shallow flow equations derived from mass and momentum balance. Examples are TITAN2D \citep{Pitman2003}, DAN3D \citep{Hungr2009}, RAMMS \citep{Christen2010}, r.avaflow \citep{Mergili2017}, faSavageHutterFOAM \citep{Rauter2018}, etc., see \citet{McDougall2017} for a review. 

Such models generally require a variety of input data, including release area and volume (a release polygon given as a shape file or a raster map of release heights), flow resistance parameters (dry-Coulomb friction and turbulent friction parameters for the Voellmy rheology), and topographic data (a digital elevation model). If the input data are accurate, the models can be deterministically run to predict characteristics of the landslide's bulk behavior, such as run-out distance, impact area, spatio-temporally resolved flow height and velocity. In practice, however, the input data usually involve large uncertainties \citep{Dalbey2008}. For example, release areas and volumes of landslides are challenging to predict due to the complexity of geological pre-conditioning factors and often a lack in subsurface information. They may be approximated by heavily-tailed probability density functions based on the statistical properties of landslide inventories \citep{Luna2013}. The flow resistance parameters are more conceptual than physical \citep{Fischer2015}. They are generally obtained by back-analyzing past events. The choice of flow resistance parameters for landslide run-out forecasting thus naturally involves uncertainties. Topographic data may also be subject to uncertainties due to error introduced during source data acquisition or data processing \citep{Zhao2020}. Therefore, it is essential to study the model's sensitivity to uncertain inputs, which could improve our understanding of the computational landslide run-out models and provide guidelines for their future usage.  

Sensitivity analyses on landslide run-out models are commonly based upon local one-at-a-time approaches, i.e., changing one input variable at a time while keeping others at their baseline values to explore its isolated effect on model outputs \citep{Hussin2012,Schraml2015}. While straightforward to implement, these types of local sensitivity analysis methods cannot assess potential interactions between input variables. Their result may highly depend on the chosen baseline values \citep{Girard2016}. In contrast, variance-based global sensitivity analyses can fully explore the input space, quantify the contribution of each variable to the output variation, and identify interactions between different variables. The Sobol' method, one typical variance-based method, has been developed and widely used since 1990s \citep{Sobol1993, Sobol2001, Saltelli2002, Saltelli2010}. The calculation of Sobol' sensitivity indices usually requires Monte Carlo-based methods, leading to a large number of necessary model evaluations. For computationally demanding models, the calculation may be prohibitively expensive. In that case, it is rather promising to employ emulation techniques to overcome the computational challenge.

An emulator is a statistical representation of a computationally demanding model referred to as a simulator. While it comes at the prize of an additional statistical error, it is typically evaluated several orders of magnitude faster than the simulator. Emulation techniques have been developed since 1980s \citep{Currin1991, OHagan2006}. Many researchers have utilized them for the purpose of global sensitivity analyses in different fields \citep{Lee2011, Rohmer2011, Lee2012, Bounceur2015, Girard2016, Aleksankina2019}. These studies either focus on emulating the evaluation of a few scalar outputs \citep{Lee2011, Rohmer2011, Girard2016}, or build separate emulators for each of the many outputs \citep{Lee2012, Aleksankina2019}. One exception among them is \citet{Bounceur2015}, who combine emulation techniques with the principal component analysis leading to emulation of a reduced-order model. For a simulator with massive outputs like a landslide run-out model, building separate emulators for each output can be computationally intensive \citep{Gu2016}. In recent years, great improvement has been made to enable simultaneous emulation for multi-output models, see for instance \citet{Rougier2008} and \citet{Gu2016}.

The goal of this study is twofold: The first is a methodological goal, namely to combine the recent development of emulation techniques \citep{Gu2016, Gu2018, Gu2019}, landslide run-out models \citep{Mergili2017}, and global sensitivity analyses \citep{LeGratiet2014} to enable global sensitivity analyses of computationally demanding landslide run-out models for the first time. The second goal is application-oriented and aims at employing the methodology to assess the relative importance of different uncertain inputs, specifically flow resistance parameters and the release volume, and their interactions in landslide run-out models based on the 2017 Bondo landslide event as a test case. 

This paper is set out as follows. In section \ref{S:2} the methodology is described, including the computational landslide run-out model based on the Voellmy rheology, Sobol' sensitivity analysis, Gaussian process (GP) emulation, and an algorithm to take emulator uncertainty into account. Section \ref{S:3} presents our Python-based implementation. Section \ref{S:4} describes the case study. Section \ref{S:5} is devoted to a discussion of our results. In section \ref{S:6}, important conclusions are drawn.

%% 2. Methodology
\section{Methodology}
\label{S:2}

\subsection{Computational landslide run-out model based on the Voellmy rheology}
\label{S:2.1}
Depth-averaged shallow flow type process models have gained popularity in practice and in academia, owing to their good compromise between accuracy and computing time \citep{Rauter2018}. A variety of flow resistance laws can be used with the models depending on landslide types and characteristics of flow material \citep{Naef2006, Hungr2009}. In case of flow-like landslides, the Voellmy rheology is one of the most widely used flow resistance laws \citep{Hussin2012, Frank2015, Schraml2015, Bevilacqua2019}. The governing system of the depth-averaged model employing the Voellmy rheology can be expressed in a surface-induced coordinate system as \citep{Christen2010, Fischer2012}
% vsmodel
\begin{linenomath}
\begin{equation}\label{eq:vsmodel}
\scalebox{.75}{$\displaystyle
\frac{\partial}{\partial_t}
\begin{pmatrix} 
h\\ hu_X\\ hu_Y 
\end{pmatrix} + \frac{\partial}{\partial_X}
\begin{pmatrix} 
hu_X\\ hu^{2}_{X} + g_Z k_{a/p} \frac{h^2}{2}\\ hu_Xu_Y \end{pmatrix} + \frac{\partial}{\partial_Y}
\begin{pmatrix} 
hu_Y\\ hu_Xu_Y \\ hu^{2}_{Y} + g_Z k_{a/p} \frac{h^2}{2}
\end{pmatrix} = \\
\begin{pmatrix}
0 \\
g_Xh - \frac{u_X}{\left \| \mathbf{u} \right\|} (\mu g_Zh + \frac{g}{\xi}{ \left \| \mathbf{u} \right\|}^2 ) \\
g_Yh - \frac{u_Y}{\left \| \mathbf{u} \right\|} (\mu g_Zh + \frac{g}{\xi}{ \left \| \mathbf{u} \right\|}^2 )
\end{pmatrix} $},
\end{equation}
\end{linenomath}
where $X,Y,Z$ denote coordinates in the down-slope, cross-slope and normal directions; $t$ denotes time; $h$ represents flow height; $u_X$ and $u_Y$ represent components of the depth-averaged surface tangent flow velocity $\mathbf{u}$ along $X$ and $Y$ directions; $g_X,g_Y,g_Z$ are components of the gravitational acceleration; $\mu$ and $\xi$ are the dry-Coulomb friction coefficient and turbulent friction coefficient, which describe the flow resistance law known as the Voellmy rheology. 

The process model is solved forward in time, hence an initial condition $h(X,Y,t_0)$ and $\mathbf{u}(X,Y,t_0)$ is needed. Typically $\mathbf{u}(X,Y,t_0)$ is zero and $h(X,Y,t_0)$ denotes the release volume and release area. Other essential inputs include the flow resistance parameters and a digital elevation map of the topography. As stated in the introduction, these input data usually involve uncertainties. The uncertainty of topographic data may be reduced by using high accuracy remote sensing data. The uncertainty of the release volume and release area may be more difficult to constrain due to the complexity of geological pre-conditioning factors and often a lack in subsurface information. This is often based on expert judgement. The flow resistance parameters depend on back-analyzing past events. It is still a great challenge to select them for quantitative risk assessment in practice \citep{McDougall2017}. In this study, we focus on the sensitivity of selected model outputs to the release volume $v_0$ (denoting the landslide magnitude) and the two flow resistance parameters $\mu$ and $\xi$ of the Voellmy rheology.

The process model produces numerous outputs, essentially given by flow height $h$ and flow velocity $\mathbf{u}$ at every space-time grid point. For the purpose of hazard assessment and mitigation, maximum values over the time duration are most interesting. In addition, aggregated scalar outputs like the apparent friction angle or impact area are commonly used to indicate the overall landslide impact. In this study, the following model outputs are under investigation.

\begin{enumerate}[$\bullet$]
\item Apparent friction angle, the tangent of which equals to the ratio of the landslide fall height and run-out distance \citep{DeBlasio2008}. It generally decreases as the run-out distance increases. 
\item Impact area, defined as the area of the region where maximum flow height values exceed a threshold value, here 0.1 $m$.
\item Maximum flow height over time at $k$ locations $\{(X_j,Y_j)\}_{j=1,\ldots,k}$, denoted as $(h_{l_1}^{\text{max}}$,\ldots,$h_{l_k}^{\text{max}})^T$. 
\item Maximum flow velocity over time at $k$ locations $\{(X_j,Y_j)\}_{j=1,\ldots,k}$, denoted as $(\|\mathbf{u}_{l_1}\|^{\text{max}}$,\ldots,$\|\mathbf{u}_{l_k}\|^{\text{max}})^T$.
\end{enumerate}

\subsection{Sobol' sensitivity analysis}
\label{S:2.2}
Assume that a simulator is denoted by $f(\mathbf{x})$ with a $p$-dimensional input $\mathbf{x}=(x_1,\ldots,x_p)^T \in {\mathbb{R}}^p$ and a scalar output $y \in \mathbb{R}$. For the process model described in section \ref{S:2.1}, $\mathbf{x}$ is a three-dimensional vector consisting of the two friction coefficients and the release volume, namely $\mathbf{x}=(\mu,\xi,v_0)^T$; $y$ could be an aggregated scalar output like apparent friction angle or impact area, or an element of a vector output like maximum flow height or velocity at a specific location. Input uncertainties of $\mathbf{x}$ induce output uncertainty of $y$. The essential idea of a Sobol' sensitivity analysis is to decompose the variance of $y$ into contributions caused by each $x_i$ and their interactions. In practice, $p$ first-order indices $\{S_i\}_{i=1,\ldots,p}$ and $p$ total-effect indices $\{S_{Ti}\}_{i=1,\ldots,p}$ are usually computed. They are defined as \citep{Saltelli2010} 
% sobol indices
\begin{linenomath}
\begin{subequations}
\begin{gather}
S_i =\frac{V_{x_i} ( E_{\mathbf{x}_{-i}}(y|x_i) )}{V(y)},\label{eq:Si}\\
S_{Ti} =1 - \frac{ V_{\mathbf{x}_{-i}} ( E_{x_i}(y|\mathbf{x}_{-i}) )}{V(y)}, \label{eq:STi}
\end{gather}
\end{subequations}
\end{linenomath}
where $V$ and $E$ represent the variance and expectation operator respectively, $\mathbf{x}_{-i}$ denotes the vector consisting of all input factors except  $x_i$. A first-order index $S_i$ accounts for the contribution of the input factor $x_i$ to the variance of the output, independent from other input factors $\mathbf{x}_{-i}$; a total-effect index $S_{Ti}$ indicates the total contribution of $x_i$ to the output variation, i.e. the sum of its first-order contribution and all high-order effects owing to interactions \citep{Saltelli2008}. The difference $S_{Ti}-S_i$ thus indicates any interaction between $x_i$ and $\mathbf{x}_{-i}$. Employing this concept to landslide run-out models will hence allow us to investigate combined effects of the two friction coefficients and the release volume on simulation outputs.

Computing the conditional variances in eqs.~(\ref{eq:Si})-(\ref{eq:STi}) involves nested integrals \citep{Girard2016}. This is analytically impractical for complex simulators like landslide run-out models. Instead, Monte Carlo-based methods are commonly used to estimate the Sobol' indices. The uncertainty introduced by Monte Carlo-based integration can be taken into account using a bootstrap strategy \citep{Archer1997}. 

In this study, we employ the numerical procedure presented in \citet{Saltelli2010}. The computational cost is $N\cdot(p+2)$ evaluations of a simulator, where $N$ is the base sample size. More specifically, the denominator $V(y)$ in eqs.~(\ref{eq:Si})-(\ref{eq:STi}) can be estimated using $2\cdot N$ simulation runs based on two independent sets of input samples. Each set consists of $N$ input samples for the simulator. Moreover, each pair of numerators in eqs.~(\ref{eq:Si})-(\ref{eq:STi}) requires additional $N$ simulation runs corresponding to a new set of $N$ input samples, which is constructed from the two independent sets. It leads to additional $p\cdot N$ simulation runs. For the detailed procedure, please refer to \citet{Saltelli2010}.

As pointed out in \citet{Saltelli2010}, $N$ should be sufficiently large, e.g. 500 or higher, which is critical in our case as the landslide run-out model itself is computationally intensive. If a single run of the simulator described in section \ref{S:2.1} costs 32 minutes, which corresponds to the average run time of the 200 simulation runs in section \ref{S:4.3}, the sensitivity analysis for three input variables will cost at least $32 \times 500 \times (3+2)=80000$ minutes, roughly 56 days. Therefore, it is necessary to employ emulation techniques to improve the computational efficiency in order to carry out this type of global sensitivity analysis.

\subsection{Gaussian process emulation}
\label{S:2.3}
A simulator, such as the landslide run-out model, represents a deterministic input-output mapping. It is usually computationally impractical to directly use such simulator for analysis requiring a large number of simulation runs, e.g. a global sensitivity analysis described in the previous section, or an uncertainty quantification, or a model calibration. In that case, GP emulators have been widely employed owing to their robustness and rich theoretical background \citep{Girard2016}. GP emulation views a simulator as an unknown function from a Bayesian perspective; the prior belief of the simulator behavior, namely a Gaussian process, is updated based on a modest number of simulation runs, leading to a posterior which can be evaluated much faster than the simulator and can then be used for computationally demanding analyses. The fundamental assumption of GP emulation is that the simulator is a smooth continuous function of its inputs \citep{OHagan2006}. Here, we recap the principal ideas of GP emulators used in this study, for detailed information please refer to \citet{OHagan1994, Bastos2009, Gu2016, Gu2018}. 

\subsubsection{Gaussian process emulator for a scalar output}
\label{S:2.3.1}
Let $f(\mathbf{x})$ denote a simulator with a $p$-dimensional input $\mathbf{x}=(x_1,\ldots,x_p)^T \in {\mathbb{R}}^p$ and a scalar output $y \in \mathbb{R}$. For example, if $f(\mathbf{x})$ is the landslide run-out model, $\mathbf{x}$ is the triplet consisting of the release volume and the two friction coefficients, and $y$ is the apparent friction angle or impact area. $f(\mathbf{x})$ is regarded as an unknown function and will be modeled as a Gaussian process. The Gaussian process is defined by a mean function $m(\cdot)$ and a covariance function $\sigma^2 c(\cdot,\cdot)$ with variance $\sigma^2$ and correlation function $c(\cdot,\cdot)$, hence
% gaussian process
\begin{linenomath}
\begin{equation}\label{eq:GP}
f(\cdot) \sim \mathcal{GP}(m(\cdot), \sigma^2 c(\cdot,\cdot)).
\end{equation}
\end{linenomath}
The mean function for any input $\mathbf{x}$ is given by the regression
% mean function
\begin{linenomath}
\begin{equation}\label{eq:GPmean}
m(\mathbf{x})=\mathbf{h}^T(\mathbf{x}) \boldsymbol{\theta},
\end{equation}
\end{linenomath}
where $\mathbf{h}(\mathbf{x})=\left(h_{1}(\mathbf{x}), h_{2}(\mathbf{x}), \ldots, h_{q}(\mathbf{x})\right)^T$ is a $q$-dimensional vector specifying basis functions, e.g. $\mathbf{h}(\mathbf{x})=(1,x_1,\ldots,x_p)^T$ for a simple linear regression, and $\boldsymbol{\theta}=\left(\theta_{1}, \theta_{2}, \ldots, \theta_{q}\right)^T$ is the corresponding $q$-dimensional vector consisting of $q$ unknown regression parameters. There are a variety of choices for the correlation functions like power exponentials, sphericals, Mat\'{e}rn, etc. The Mat\'{e}rn correlation function is chosen here following \citet{Gu2018}. For any $\mathbf{x}_i=(x_{i1},\ldots,x_{ip})^T$ and $\mathbf{x}_j=(x_{j1},\ldots,x_{jp})^T$, their correlation is described by
% correlation function
\begin{linenomath}
\begin{equation}\label{eq:GPcorrelation}
c\left(\mathbf{x}_{i}, \mathbf{x}_{j}\right) = \prod_{l=1}^{p} \left(1+\frac{\sqrt{5}d_l}{\gamma_l}+\frac{5d_{l}^2}{3\gamma_{l}^{2}} \right) \exp{\left(-\frac{\sqrt{5}d_l}{\gamma_l}\right)},
\end{equation}
\end{linenomath}
where $d_l=|x_{il}-x_{jl}|$ represents the distance between the two inputs in the $l$-th dimension, and $\boldsymbol{\gamma}=(\gamma_1,\ldots,\gamma_p)^T$ is a $p$-dimensional vector consisting of $p$ unknown range parameters.

Eqs.~(\ref{eq:GP})-(\ref{eq:GPcorrelation}) represent the prior belief of the simulator's behavior. The fundamental idea now is to update the prior belief following a Bayesian methodology based on evaluations of the simulator at $N_{sim}$ selected inputs $\mathbf{x}^{\mathcal{D}}=\{\mathbf{x}_i\}_{i=1,\ldots,N_{sim}}$. Owing to the property of the Gaussian process, the outputs corresponding to $\mathbf{x}^{\mathcal{D}}$, denoted as $\mathbf{y}^{\mathcal{D}}=\{f(\mathbf{x}_i)\}_{i=1,\ldots,N_{sim}}$, follow a multivariate Gaussian distribution
% likelihood
\begin{linenomath}
\begin{equation} \label{eq:likelihood}
\mathbf{y}^{\mathcal{D}} | \boldsymbol{\theta}, \sigma^{2}, \boldsymbol{\gamma}  \sim \mathcal{N}_{N_{sim}}\left(\mathbf{H}\boldsymbol{\theta}, \sigma^{2} \mathbf{R}\right),
\end{equation}
\end{linenomath}
where $\mathbf{H}=\left[\mathbf{h}(\mathbf{x}_{1}), \ldots, \mathbf{h}(\mathbf{x}_{N_{sim}}) \right]^T$ is the $N_{sim} \times q$ basis design matrix and $\mathbf{R}$ is the $N_{sim} \times N_{sim}$ correlation matrix with $(i,j)$ element $c(\mathbf{x}_{i}, \mathbf{x}_{j})$. Again, owing to the property of the Gaussian process, the output $y^*$ at any new input $\mathbf{x}^*$ follows a Gaussian distribution conditioned on $\mathbf{y}^{\mathcal{D}}$, given by
% conditional Gaussian
\begin{linenomath}
\begin{subequations}
\begin{gather}
y^* | \mathbf{y}^{\mathcal{D}}, \boldsymbol{\theta}, \sigma^{2}, \boldsymbol{\gamma} \sim  \mathcal{N}
\left(m', \sigma^2 c' \right),\label{eq:conditionalGaussian}\\
m'=\mathbf{h}^T(\mathbf{x}^*) \boldsymbol{\theta} + \mathbf{r}^T(\mathbf{x}^*) \mathbf{R}^{-1} (\mathbf{y}^{\mathcal{D}}-\mathbf{H}\boldsymbol{\theta}), \label{eq:conditionalGaussianM}\\
c'=c(\mathbf{x}^*, \mathbf{x}^*) - \mathbf{r}^T(\mathbf{x}^*) \mathbf{R}^{-1} \mathbf{r}(\mathbf{x}^*), \label{eq:conditionalGaussianC}
\end{gather}
\end{subequations}
\end{linenomath}
where $\mathbf{r}(\mathbf{x}^*)=\left(c(\mathbf{x}^*, \mathbf{x}_{1}), \ldots, c(\mathbf{x}^*, \mathbf{x}_{N_{sim}}) \right)^T$.

The parameters $\boldsymbol{\theta}$, $\sigma^2$, and $\boldsymbol{\gamma}$ in eq.~(\ref{eq:conditionalGaussian}) are the unknowns that need to be updated. Of these, regression parameters $\boldsymbol{\theta}$ and the variance $\sigma^2$ can be integrated out using a conjugate analysis and Bayes' theorem. More specifically, a weak prior for $(\boldsymbol{\theta},\sigma^2)$ is assumed to have the form $p(\boldsymbol{\theta},\sigma^2) \propto (\sigma^2)^{-1}$, which is within the conjugate family as the likelihood, i.e. eq.~(\ref{eq:likelihood}). Combining the weak prior and the likelihood gives the posterior $p(\boldsymbol{\theta},\sigma^2|\mathbf{y}^{\mathcal{D}},\boldsymbol{\gamma})$. Then, $\boldsymbol{\theta}$ and $\sigma^2$ are successively integrated out from eq.~(\ref{eq:conditionalGaussian}) by applying the Bayesian chain rule to $p(\boldsymbol{\theta},\sigma^2|\mathbf{y}^{\mathcal{D}},\boldsymbol{\gamma})$ and eq.~(\ref{eq:conditionalGaussian}). This yields a Student's t-distribution with ${N_{sim}}-q$ degrees of freedom, which describes the distribution of $y^*$ conditioned on $\mathbf{y}^{\mathcal{D}}$ and $\boldsymbol{\gamma}$, i.e.
% t-distribution
\begin{linenomath}
\begin{subequations}
\begin{gather}
y^* | \mathbf{y}^{\mathcal{D}}, \boldsymbol{\gamma}
\sim \mathcal{S}t(m'', {\hat{\sigma}}^2c'',{N_{sim}}-q), \label{eq:student} \\
m'' = \mathbf{h}^T(\mathbf{x}^*) \hat{\boldsymbol{\theta}} + \mathbf{r}^T(\mathbf{x}^*) \mathbf{R}^{-1} (\mathbf{y}^\mathcal{D}-\mathbf{H}\hat{\boldsymbol{\theta}}), \label{eq:studentM}\\
{\hat{\sigma}}^2=(N_{sim}-q)^{-1}(\mathbf{y}^\mathcal{D}-\mathbf{H}\hat{\boldsymbol{\theta}})^T \mathbf{R}^{-1} (\mathbf{y}^\mathcal{D}-\mathbf{H}\hat{\boldsymbol{\theta}}), \label{eq:studentS}\\
\begin{split}
c''=  c(\mathbf{x}^*, \mathbf{x}^*) - \mathbf{r}^T(\mathbf{x}^*) \mathbf{R}^{-1} \mathbf{r}(\mathbf{x}^*) + 
\left(\mathbf{r}^T(\mathbf{x}^*) \mathbf{R}^{-1} \mathbf{H}-\mathbf{h}^T(\mathbf{x}^*)\right) \\
\times (\mathbf{H}^T\mathbf{R}^{-1}\mathbf{H})^{-1}
\left(\mathbf{r}^T(\mathbf{x}^*) \mathbf{R}^{-1} \mathbf{H}-\mathbf{h}^T(\mathbf{x}^*)\right)^T, \label{eq:studentC}\end{split}
\end{gather}
\end{subequations}
\end{linenomath}
where $\hat{\boldsymbol{\theta}}=(\mathbf{H}^T\mathbf{R}^{-1}\mathbf{H})^{-1}\mathbf{H}^T\mathbf{R}^{-1}\mathbf{y}^{\mathcal{D}}$. From a Bayesian viewpoint, the remaining unknown $\boldsymbol{\gamma}$ in eq.~(\ref{eq:student}) should also be integrated out by employing a certain prior for $\boldsymbol{\gamma}$. The integral, however, is highly intractable and would require computationally intensive methods like Markov Chain Monte Carlo sampling strategies. Instead, $\boldsymbol{\gamma}$ is often estimated by solving an optimization problem, e.g. maximizing its marginal likelihood or finding its marginal posterior mode. In this study, we use the marginal posterior mode estimation, recommended by \citet{Gu2018} due to its robustness. Substituting the marginal posterior mode estimation of $\boldsymbol{\gamma}$ into eqs.~(\ref{eq:student})-(\ref{eq:studentC}), finally, gives the GP emulator, denoted as $\hat{f}(\mathbf{x})$. It provides a prediction of the simulator output at any new input $\mathbf{x}^*$ in the form of eq.~(\ref{eq:studentM}), as well as an assessment of the prediction uncertainty, like a 95\% credible interval (CI(95\%)) of the prediction.  

\subsubsection{Gaussian process emulator for a vector output}
\label{S:2.3.2}
Let $\mathbf{f}(\mathbf{x})$ denote a simulator with a $p$-dimensional input $\mathbf{x}=(x_1,\ldots,x_p)^T \in {\mathbb{R}}^p$ and a $k$-dimensional output $\mathbf{y}=(y_1,\ldots,y_k)^T \in {\mathbb{R}}^k$. For example, $\mathbf{f}(\mathbf{x})$ is the landslide run-out model, $\mathbf{x}$ is the triplet consisting of the release volume and the two flow resistance parameters, and $\mathbf{y}$ is maximum flow height or velocity over time at $k$ locations. In a straightforward Many Single emulator approach \citep{Gu2016}, each component of the simulator, i.e. $\{y_j=f_j(\mathbf{x})\}_{j=1,\ldots,k}$, is assumed to follow an independent Gaussian process having the form of eq.~(\ref{eq:GP}), with independent parameters $\{\boldsymbol{\theta}_j\}_{j=1,\ldots,k}$, $\{\sigma_{j}^{2}\}_{j=1,\ldots,k}$, and $\{\boldsymbol{\gamma}_j\}_{j=1,\ldots,k}$. For each independent emulator, the range parameters $\boldsymbol{\gamma}_j=(\gamma_{j1},\ldots,\gamma_{jp})^T$ need to be estimated by solving an optimization problem as described in section \ref{S:2.3.1}. As a consequence, the training of the emulators may take  a lot of time when $k$ is large.

In this study, we use however an alternative approach, namely the parallel partial GP emulator developed by \citet{Gu2016} to simultaneously emulate the relation between the $p$-dimensional input and $k$-dimensional output. Similar to the Many Single emulator approach, each element of the simulator is assumed to follow an independent Gaussian process of the form eq.~(\ref{eq:GP}). The main difference is that all of the $k$ Gaussian processes are assumed to share common range parameters $\boldsymbol{\gamma}$, which are then estimated from the overall likelihood \citep{Gu2016}. The $q$-dimensional basis functions $\mathbf{h}(\mathbf{x})=\left(h_{1}(\mathbf{x}), h_{2}(\mathbf{x}), \ldots, h_{q}(\mathbf{x})\right)^T$ are also assumed to be the same. These modifications greatly reduce the emulator training time. Once the estimation of the common $\boldsymbol{\gamma}$ is obtained, the parallel partial GP emulator is determined, which is now a collection of $k$ Student's t-distributions. Here, it is denoted as $\{\hat{f}_j(\mathbf{x})\}_{j=1,\ldots,k}$. The exact form of the emulator can be found in \citet{Gu2016}.

\subsection{Emulator uncertainty in Sobol' sensitivity analysis}
\label{S:2.4}
The efficiency improvement by using GP emulators comes at a cost, i.e. additional emulator uncertainty. We can quantify this type of uncertainty as it can be evaluated from the emulator directly. Yet, we need to find a way to account for this uncertainty in the subsequent analysis. Alongside the development of emulation techniques and global sensitivity analysis methods, a number of approaches have been developed in recent years to address this issue in global sensitivity analyses, e.g. \citet{Oakely2004, Marrel2009, Janon2014, LeGratiet2014}. 

% Algorithm
\begin{algorithm}[hp]
    \caption{Emulator-based Sobol' index evaluation}
  \begin{algorithmic}[1]
    \STATE Choose input configurations $\mathbf{x}^{\mathcal{D}}=\{\mathbf{x}_i\}_{i=1,\ldots,N_{sim}}$, usually using a Latin hypercube design. \label{al:step1}
    \STATE Run the simulator $f({\mathbf{x}})$, e.g. the landslide run-out model, at each of $N_{sim}$ chosen input configurations to obtain outputs $\mathbf{y}^{\mathcal{D}}$ (see section \ref{S:2.1}). \label{al:step2}
    \STATE Build the emulator $\hat{f}({\mathbf{x}})$ based on $\mathbf{x}^{\mathcal{D}}$-$\mathbf{y}^{\mathcal{D}}$ (see section \ref{S:2.3.1}). \label{al:step3}
    \STATE Set the base sample size $N$, realization sample size $N_r$, and bootstrap sample size $N_b$. Sample inputs $\{\mathbf{x}_i\}_{i=1,\ldots,N \cdot (p+2)}$ from the input domain according to \citet{Saltelli2010} (see section \ref{S:2.2}).
    \FOR{$n_r =1,\ldots,N_r$} \label{al:step5}
      \STATE \begin{varwidth}[t]{\linewidth}
      Sample a set of $N \cdot (p+2)$ realizations of $\hat{f}(\mathbf{x})$ with $\{\mathbf{x}_i\}_{i=1,\ldots,N \cdot (p+2)}$, \par
      denoted as $\{\hat{f}^{n_r}(\mathbf{x}_i)\}_{i=1,\ldots,N \cdot (p+2)}$.
       \end{varwidth}
      \STATE \begin{varwidth}[t]{\linewidth} 
      Compute $\{\hat{S}_{i}^{n_r,1}\}_{i=1,\ldots,p}$ and $\{\hat{S}_{Ti}^{n_r,1}\}_{i=1,\ldots,p}$ based on the realizations  \par
      $\{\hat{f}^{n_r}(\mathbf{x}_i)\}_{i=1,\ldots,N \cdot (p+2)}$.
       \end{varwidth}
      \FOR{$n_b =2,\ldots,N_b$}
        \STATE \begin{varwidth}[t]{\linewidth} 
        Sample with replacements $\{\tilde{\mathbf{x}}_i\}_{i=1,\ldots,N \cdot (p+2)}$ from $\{\mathbf{x}_i\}_{i=1,\ldots,N \cdot (p+2)}$ \par
        and record realizations $\{\hat{f}^{n_r}(\tilde{\mathbf{x}}_i)\}_{i=1,\ldots,N \cdot (p+2)}$.
         \end{varwidth}
        \STATE \begin{varwidth}[t]{\linewidth} 
        Compute $\{\hat{S}_{i}^{n_r,n_b}\}_{i=1,\ldots,p}$ and $\{\hat{S}_{Ti}^{n_r,n_b}\}_{i=1,\ldots,p}$ based on the realizations 
        \hskip\algorithmicindent $\{\hat{f}^{n_r}(\tilde{\mathbf{x}}_i)\}_{i=1,\ldots,N \cdot (p+2)}$.
         \end{varwidth}
      \ENDFOR
    \ENDFOR
    \STATE \textbf{return} $\{\hat{S}_{i}^{n_r,n_b}\}_{i=1,\ldots,p}^{n_r=1,\ldots,N_r;n_b=1,\ldots,N_b}$ and $\{\hat{S}_{Ti}^{n_r,n_b}\}_{i=1,\ldots,p}^{n_r=1,\ldots,N_r;n_b=1,\ldots,N_b}$.
    \STATE Estimate $S_i$ and $S_{Ti}$ defined in eqs.~(\ref{eq:Si})-(\ref{eq:STi}) using $\hat{S}_i=\frac{1}{N_r \cdot N_b} \sum {\hat{S}_{i}^{n_r,n_b}}$ and $\hat{S}_{Ti}=\frac{1}{N_r \cdot N_b} \sum {\hat{S}_{Ti}^{n_r,n_b}}$, with ${i=1,\ldots,p}$. Quantify the overall uncertainty (i.e. Monte Carlo-based sampling uncertainty and emulator uncertainty) of an estimated Sobol' index using its standard deviation or CI(95\%). \label{al:step14}
  \end{algorithmic}
  \label{al:emulationSobol}
\end{algorithm}

For this study, we choose to integrate the method proposed by \citet{LeGratiet2014}, which combines the work of \citet{Oakely2004} and \citet{Janon2014}. It can simultaneously take the Monte Carlo-based sampling uncertainty (section \ref{S:2.2}) and emulator uncertainty into account when calculating the Sobol' indices. We adapt the method to combine the sampling scheme presented in \citet{Saltelli2010} and the GP emulators developed by \citet{Gu2016, Gu2018}. 

The adapted method for a simulator with a scalar output, namely $f(\mathbf{x})$, is shown in Algorithm \ref{al:emulationSobol}. For a simulator with a $k$-dimensional output, i.e. $\mathbf{f}(\mathbf{x})$, the method is essentially similar. Minor modifications are as follows.

\begin{enumerate}[$\bullet$]
\item In steps~\ref{al:step1}-\ref{al:step3}, a parallel partial GP emulator $\{\hat{f}_j(\mathbf{x})\}_{j=1,\ldots,k}$ is built  (section \ref{S:2.3.2}) instead of $\hat{f}(\mathbf{x})$.
\item Steps~\ref{al:step5}-\ref{al:step14} are repeated for each $\hat{f}_j(\mathbf{x})$ to evaluate the Sobol' indices at the $j$-th element of the $k$-dimensional output, where $j=1,\ldots,k$.
\end{enumerate}

%% 3. Implementation
\section{Implementation}
\label{S:3}
The methodology presented in section \ref{S:2} involves recent progress in different fields (i.e. landslide run-out modeling, global sensitivity analysis, and GP emulation), in which respective software solutions have been developed. In this section, we present our Python-based implementation which integrates recent open-source software in those fields to a unified framework. It serves as a wrapper to realize Algorithm \ref{al:emulationSobol} for computationally demanding landslide run-out models. The principle components of the implementation are as follows.
\begin{enumerate}[$\bullet$]
\item \textbf{Simulator}. \citet{Mergili2017} presented the open source software r.avaflow for simulation of a variety of mass flows, which relies on GRASS GIS 7. It employs a Voellmy-type model (section \ref{S:2.1}) and a multi-phase mass flow model \citep{Pudasaini2019}. Here, the former is the simulator under investigation. We implemented a Python-based wrapper to automatically prepare a batch job, run simulations, and extract outputs given the selected values of input variables $\mathbf{x}^{\mathcal{D}}$, without explicitly starting GRASS and r.avaflow.
\item \textbf{Emulator}. \citet{Gu2019} presented the R package RobustGaSP (Robust Gaussian Stochastic Process Emulation), in which they implemented the marginal posterior mode estimator for the range parameters $\boldsymbol{\gamma}$ (see section \ref{S:2.3.1}) and the parallel partial GP emulator (see section \ref{S:2.3.2}). We implemented a Python-based wrapper based on rpy2 (the Python interface to the R language) to utilize RobustGaSP within the unified Python-based framework.
\item \textbf{Emulator-based Sobol' analysis}. \citet{Herman2017} presented the Python package SALib (Sensitivity Analysis Library in Python), in which the numerical procedure of calculating the Sobol' indices for a simulator is implemented. We extended their codes to realize Algorithm \ref{al:emulationSobol} which enables emulator-based Sobol' analysis for multi-output simulators.
\end{enumerate}

%% 4. Case study
\section{Case study}
\label{S:4}
\subsection{Case background}
\label{S:4.1}
\begin{figure}[tbh]
\centering
\includegraphics[width=\textwidth]{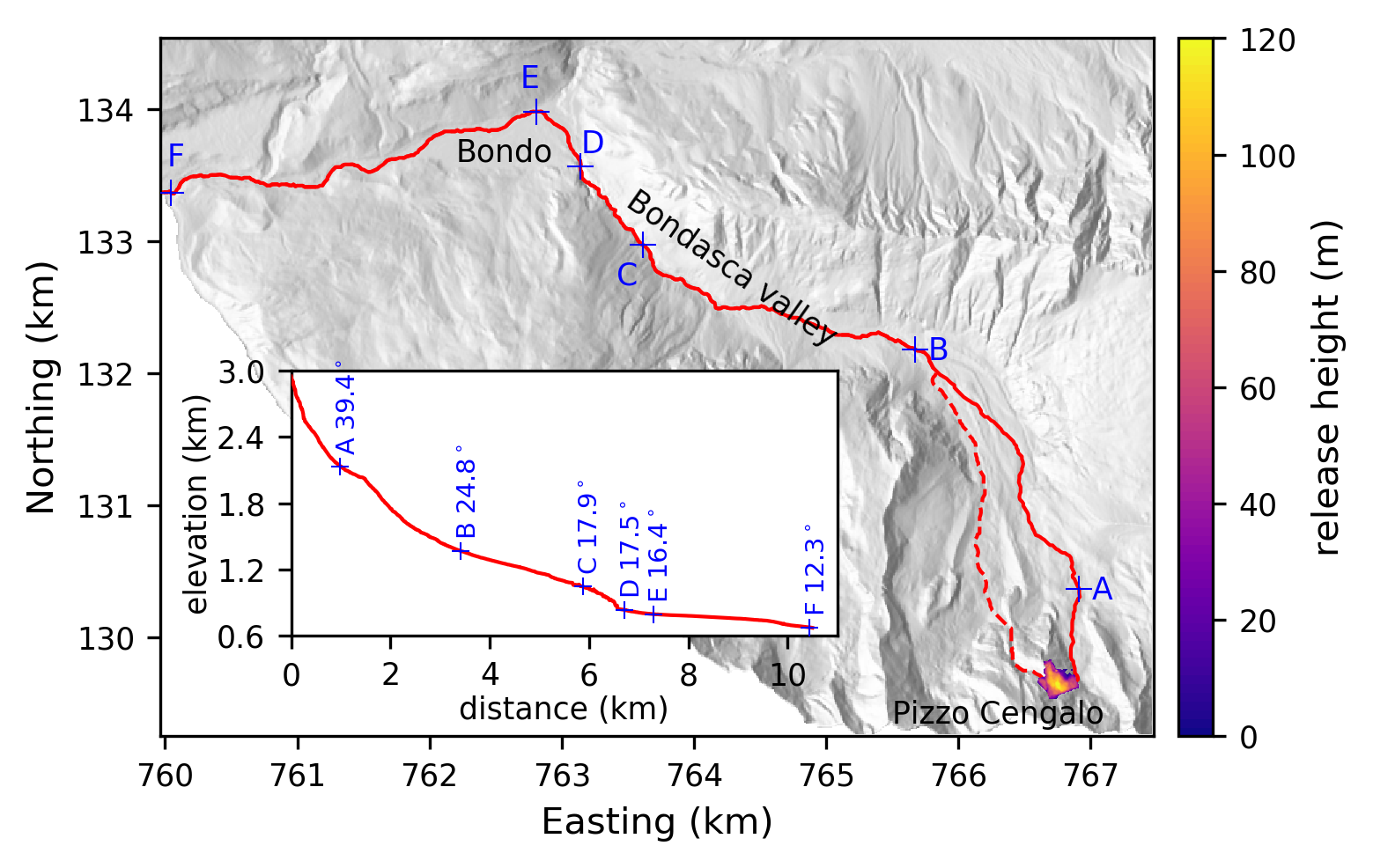}
\caption{Pizzo Cengalo-Bondo topography. The colormap shows the distribution of the release mass of the 2017 landslide event (shown in the 10 $m$ resolution computational mesh grid used for the simulations). The solid line and dashed line denote the major and minor flow paths. The embedded plot in the bottom-left corner shows the profile of the major flow path, on top of which locations A-F with respective apparent friction angles are noted for our later discussion in section \ref{S:5.2}.}
\label{fig:cengalo}
\end{figure}

Pizzo Cengalo, see figure~\ref{fig:cengalo}, located in the Swiss Alps, is subjected to rock fall and landslide events since decades due to its geological pre-conditioning factors \citep{Walter2020}. Two recent landslide events in that area are well-documented and widely studied. The first event occurred on December 27th 2011. Around 1.5 Mio $m^3$ of rock detached from the northeastern face of Pizzo Cengalo and evolved into a rock avalanche traveling 2.7 $km$ down the Bondasca valley. The second event occurred on August 23th 2017. Approximately 3 Mio $m^3$ of rock were released from the northeastern face of Pizzo Cengalo, leading to a rock avalanche traveling 3.2 $km$ down the Bondasca valley. A part of the rock avalanche turned into an initial debris flow, followed by a series of additional debris flows within 48 hours, which reached the village Bondo \citep{Walter2020}.

% figure Cengalo

Our case study is based on the topography and release area of the 2017 landslide event. A pre-event digital elevation model (DEM) and a post-event DEM are available, both with 1 $m$ resolution. They are based on airborne laser scans after the 2011 and after the 2017 events, as well as aerial images acquired by the Swiss topographic services Swisstopo \citep{Walter2020}. Release area and initial mass distribution of the event can be obtained from the height difference map of the two DEMs. As the topographic input, we use a merged DEM based on the pre-event and post-event DEMs. The merged DEM reflects the post-event topography in the release area and pre-event topography in other areas. In addition, we use the same release area as the 2017 landslide event, as shown in figure~\ref{fig:cengalo}. The grid size of the computational mesh for the simulator is set to be 10 $m$.

It should be noted, that the intention of the case study is not to back-analyze the 2017 landslide event. Other publications are devoted to that research question \citep{Mergili2020, Walter2020}. Our focus is to apply the novel emulator-based global sensitivity analysis to the Bondo event in order to assess the model's sensitivity to flow resistance parameters $\mu$ and $\xi$, as well as the release volume $v_0$ (see section \ref{S:2.1}).

\subsection{Ranges of uncertain inputs}
\label{S:4.2}
\citet{Sosio2008} summarized typical ranges for $\mu$ and $\xi$ based on a variety of literature. For rock avalanches and debris flows, the range for $\mu$ is 0.05-0.25 and that for $\xi$ is 200-1000 $m/s^2$.  \citet{Schraml2015} presented many back-analyzed $\mu$-$\xi$ sets, consisting of published values in the literature and their own case study. For most of the rock avalanche and debris flow events, $\mu$ lies within the range 0.02-0.25 and $\xi$ varies between 100-2000 $m/s^2$. \citet{Aaron2019} presented back-analyses results of a rock avalanche dataset consisting of 45 past rock avalanche events.  Their calibrated values of $\mu$ vary between 0.025-0.29, except 4 cases in which the path material is bedrock. The calibrated values of $\xi$ are in the range 200-2100 $m/s^2$. 

Based on the reference studies, we set the ranges 0.02-0.3 and 100-2200 $m/s^2$ for $\mu$ and $\xi$ respectively. As regards to the release volume $v_0$, we assume it varies between 1.5 Mio $m^3$ and 4.5 Mio $m^3$, namely $\pm50\%$ based on the 3 Mio $m^3$ release volume of the 2017 landslide event. This is achieved by multiplying the distribution of initial mass of the 2017 landslide event with a value between 0.5 and 1.5. To sum up, the three uncertain inputs result in a three dimensional input space, where $\mu$, $\xi$, and $v_0$ vary independently within 0.02-0.3, 100-2200 $m/s^2$, and 1.5-4.5 Mio $m^3$.
% figure-latin hypercube design
\begin{figure}[thpb]
\centering
\includegraphics[width=\textwidth]{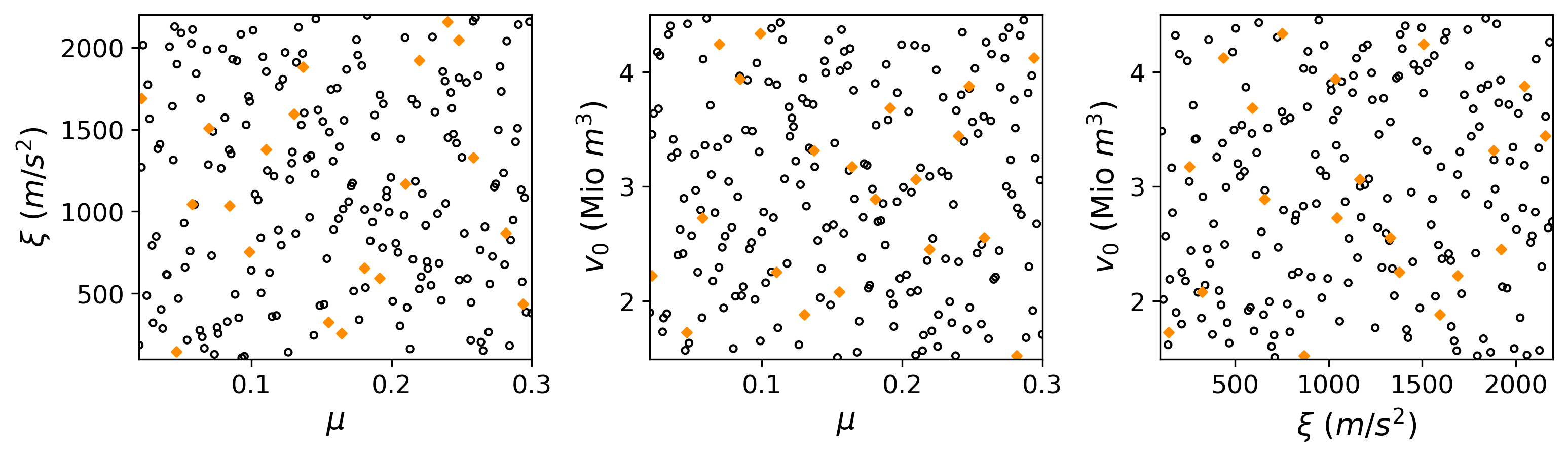}
\caption{Two-dimensional projection of the 200 training samples (void circles) and 20 validation samples (solid diamonds) from two independent maximin Latin hypercube designs. Left $\xi$-$\mu$, middle $v_0$-$\mu$, right $v_0$-$\xi$. The 200 samples are used to build the emulators. The 20 samples are used to validate the parallel partial GP emulators.}
\label{fig:lhs-samples}
\end{figure}

\subsection{Emulator design and validation}
\label{S:4.3}
To prepare the emulator training data, $N_{sim}=200$ samples are drawn from the three dimensional input space using the maximin Latin hypercube design which maximises the minimum distance between design points to achieve optimum space-filling properties \citep{Aleksankina2019}, see figure~\ref{fig:lhs-samples}. This results in $\mathbf{x}^{\mathcal{D}}=\{(\mu_i,\xi_i,v_{0i})^T\}_{i=1,\ldots,200}$. One run-out simulation takes 32 minutes on average on a laptop with Intel Core i7-9750H CPU. For each simulation run, we extract the apparent friction angle and impact area, as well as $(h_{l_1}^{\text{max}}$,\ldots,$h_{l_k}^{\text{max}})^T$ and $(\|\mathbf{u}_{l_1}\|^{\text{max}}$,\ldots,$\|\mathbf{u}_{l_k}\|^{\text{max}})^T$ at $k=47958$ chosen locations. This corresponds to the two aggregated scalar outputs and the two vector outputs in section \ref{S:2.1}. At each of the 47958 locations, at least one of the 200 simulation runs has a maximum flow height value larger than 0.1 $m$. Correspondingly, two scalar GP emulators (section \ref{S:2.3.1}) and two parallel partial GP emulators (section \ref{S:2.3.2}) are built based on $\mathbf{x}^{\mathcal{D}}$ and its respective simulation outputs. Each parallel partial GP emulator takes about 0.05 seconds to determine maximum flow height or velocity at all 47958 locations for a new input configuration.

Before using the emulators for our further sensitivity analysis, we validate their performance. The proportion of validation outputs that lie in emulator-based 95\% credible intervals is chosen as the diagnostic, denoted as $P_\text{CI(95\%)}$. This is commonly used in the literature \citep[e.g.][]{Lee2011, Spiller2014, Bounceur2015, Gu2016}. It is defined as
\begin{linenomath}
\begin{equation}\label{eq:PCI}
P_\text{CI(95\%)}=\frac{1}{n}\sum_{i=1}^{n} 1\{f(\mathbf{x}_{i}^{*}) \in \hat{f}(\mathbf{x}_{i}^{*})_\text{CI(95\%)} \},
\end{equation}
\end{linenomath}
where $n$ is the number of input configurations for validation, $f(\mathbf{x}_{i}^{*})$ and $\hat{f}(\mathbf{x}_{i}^{*})_\text{CI(95\%)}$ denote the simulation output and the CI(95\%) of the emulator prediction at the input $\mathbf{x}_{i}^{*}$ respectively. $P_\text{CI(95\%)}$ would be close to 0.95 for an ideal emulator.

% figure-scalar emulator validation
\begin{figure}[tpb]
\centering
\includegraphics[width=\textwidth]{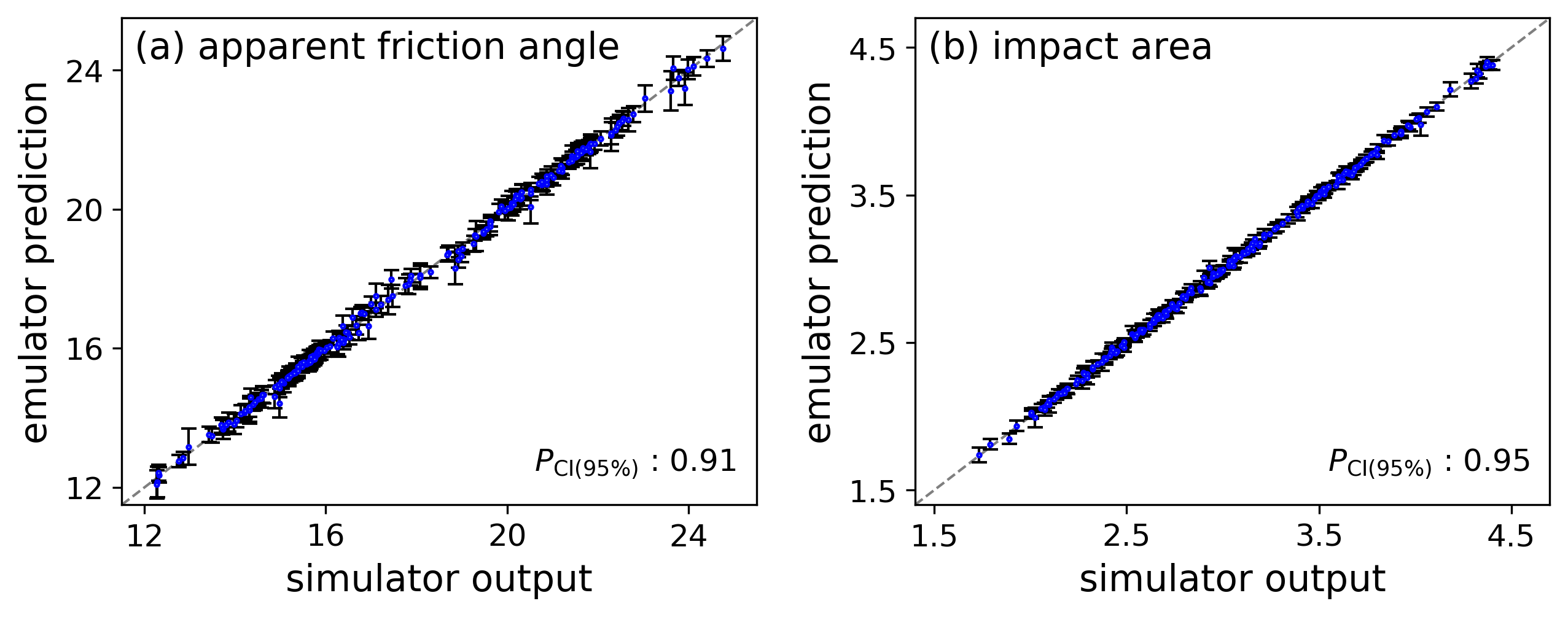}
\caption{Leave-one-out cross validation of the GP emulators for scalar outputs (a) apparent friction angle (in degree), and (b) impact area (in Mio $m^2$). The error bars denote 95\% credible intervals of the emulator predictions.}
\label{fig:scalar-validation}
\end{figure}

\begin{figure}[tpb]
\centering
\includegraphics[width=\textwidth]{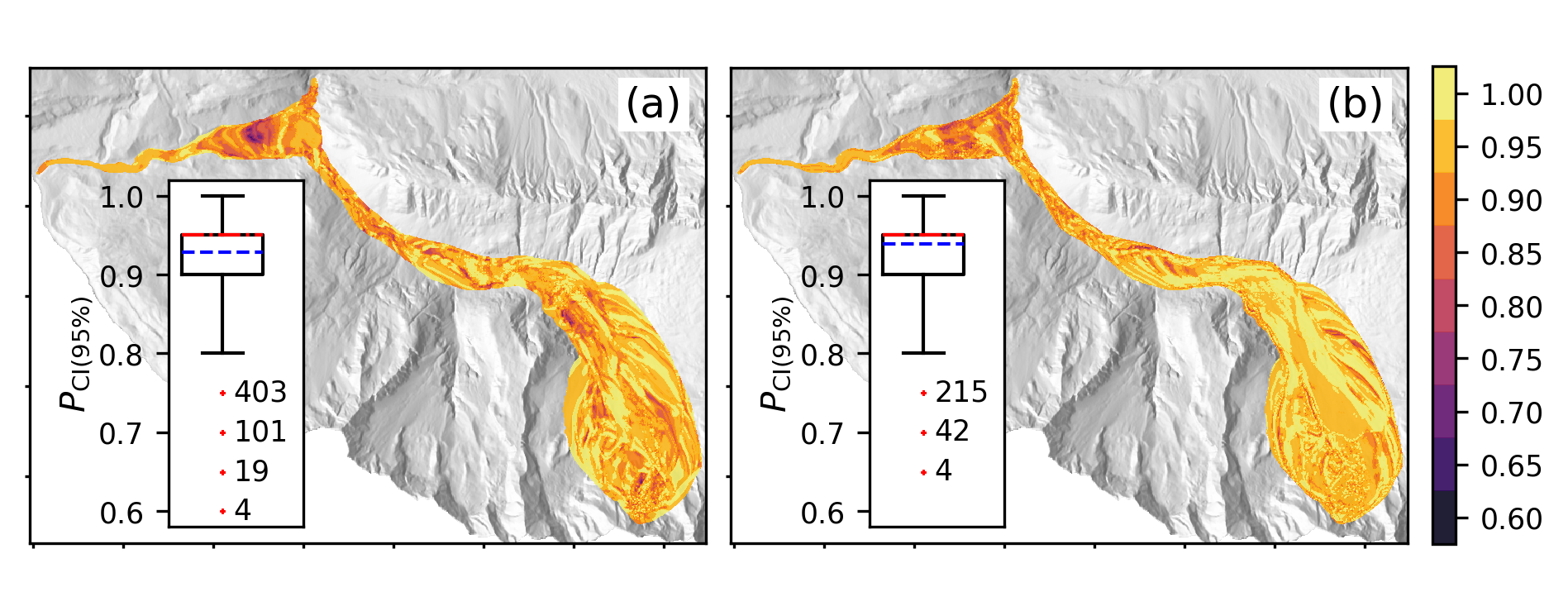}
\caption{Validation of the parallel partial GP emulators for vector outputs (a) $(h_{l_1}^{\text{max}}$,\ldots,$h_{l_k}^{\text{max}})^T$ and (b) $(\|\mathbf{u}_{l_1}\|^{\text{max}}$,\ldots,$\|\mathbf{u}_{l_k}\|^{\text{max}})^T$ with $k=47958$, using 20 validation runs based on an independent maximin Latin hypercube design. In each panel, the colormap shows the $P_\text{CI(95\%)}$ values at each location; the box plot presents the distribution of $P_\text{CI(95\%)}$ values. In the box plot, the whiskers denote the 2.5th and 97.5th percentiles; the blue dashed line denotes the mean; the number of outliers for each outlier value is noted due to overlapping. The mean of $P_\text{CI(95\%)}$ over all locations for maximum flow height/velocity is 0.93/0.94.}
\label{fig:vector-validation}
\end{figure}

The two scalar emulators are validated using the leave-one-out cross validation method as implemented in the RobustGaSP package (meaning $n=200$), see figure~\ref{fig:scalar-validation}. Both emulators perform well with emulator prediction values being close to simulator outputs and $P_\text{CI(95\%)}$ close to 0.95. As no cross validation scheme is implemented in the RobustGaSP package for a parallel partial GP emulator, we validate the two parallel partial GP emulators for $(h_{l_1}^{\text{max}}$,\ldots,$h_{l_k}^{\text{max}})^T$ and $(\|\mathbf{u}_{l_1}\|^{\text{max}}$,\ldots,$\|\mathbf{u}_{l_k}\|^{\text{max}})^T$ using additional 20 simulation runs based on an independent maximin Latin hypercube design, see figure~\ref{fig:lhs-samples}. Figure~\ref{fig:vector-validation} (a) shows $P_\text{CI(95\%)}$ values at each location and their distribution in the form of a box plot based on the maximum flow height emulator. Figure~\ref{fig:vector-validation} (b) shows the same evaluation based on the maximum flow velocity emulator. The lowest $P_\text{CI(95\%)}$ value of the maximum flow height/velocity emulator is 0.6/0.65, and 95\% of the $P_\text{CI(95\%)}$ values of both emulators are within 0.8-1. Both emulators show good performance with mean values of $P_\text{CI(95\%)}$ over all locations being 0.93 and 0.94 respectively.
% figure-vector emulator validation

\subsection{Preliminary convergence analysis}
\label{S:4.4}
\begin{figure}[h]
\centering
\includegraphics[width=\textwidth]{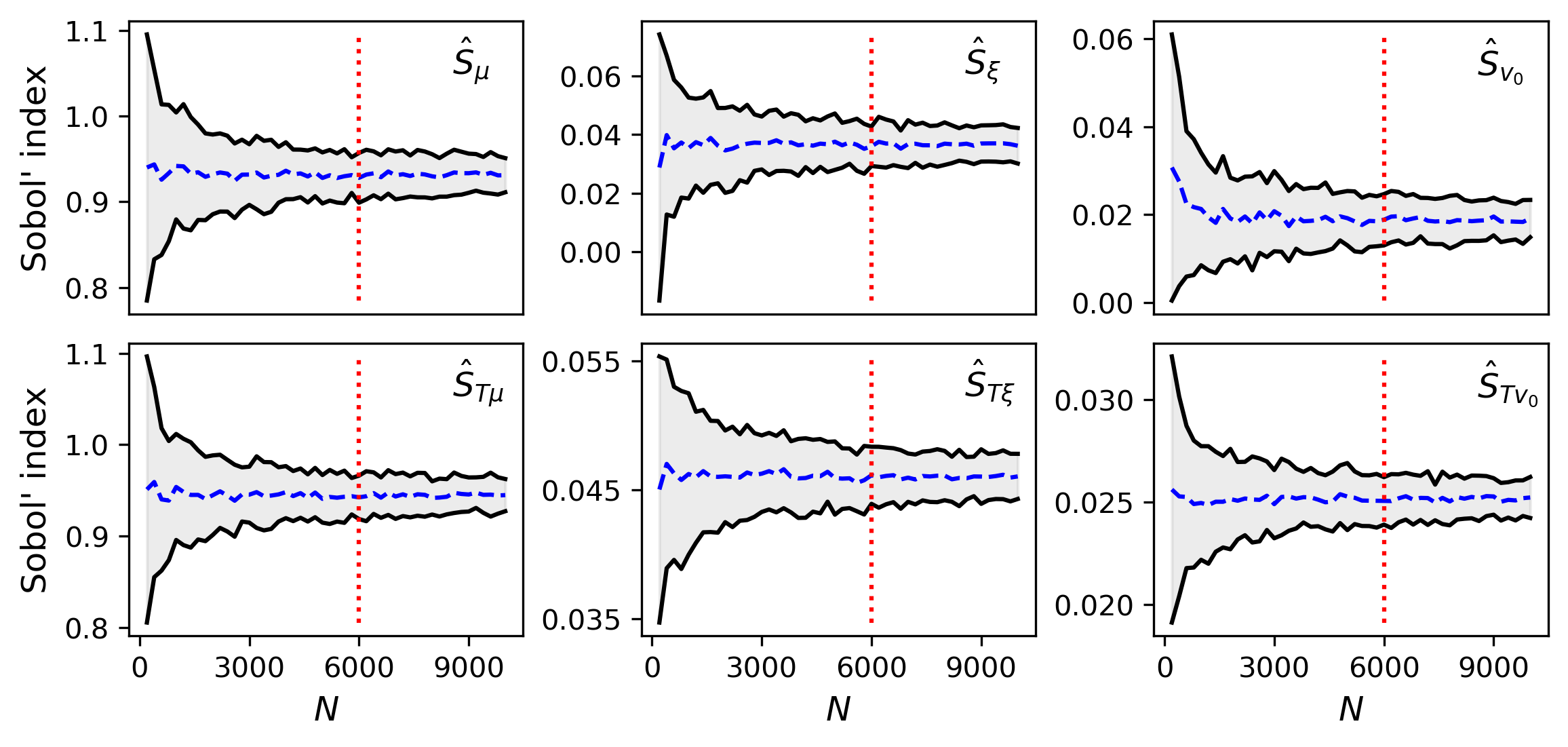}
\caption{First-order (first row) and total-effect Sobol' indices (second row) based on the GP emulator for the apparent friction angle, with $N_r=N_b=50$ and $N$ increasing from 200 to 10000 with a step size 200. In each panel, the dashed line and solid line show the change of the estimated Sobol' index and its 95\% credible interval respectively (see step \ref{al:step14} in Algorithm \ref{al:emulationSobol}); the estimated Sobol' index tends to converge for $N\geq4000$ and the length of its 95\% credible interval hardly decreases for $N\geq6000$.}
\label{fig:convergence}
\end{figure}

The base sample size $N$, realization sample size $N_r$, and bootstrap sample size $N_b$ need to be determined before using the validated emulators for the Sobol' sensitivity analysis (see Algorithm \ref{al:emulationSobol}). Here, we present the results of a convergence analysis based on the validated emulator for the apparent friction angle in order to determine values for these sample sizes. Figure~\ref{fig:convergence} shows how the estimated Sobol' indices and their CI(95\%) values change with $N$ increasing from 200 to 10000 with a step size 200, while keeping $N_r=N_b=50$. It can be seen that the estimated Sobol' indices tend to converge when $N$ is large than 4000, and their CI(95\%) lengths almost do not decrease for $N\geq6000$. We conducted the same analysis with $N_r=N_b=100$ and $N_r=N_b=200$. The results are similar to our findings with $N_r=N_b=50$, indicating little impact of $N_r$ and $N_b$. Therefore, we set $N=6000$ and $N_r=N_b=50$ for the following sensitivity study. It leads to $N \cdot (p+2)=6000 \cdot (3+2)=30000$ samples from the three dimensional input space to estimate the Sobol' indices, namely $\{(\mu_i,\xi_i,v_{0_i})^T\}_{i=1,\ldots,30000}$. Among them, $2\cdot N=12000$ samples are used to estimate the overall variance term $V(y)$ in eqs.~(\ref{eq:Si})-(\ref{eq:STi}), see section~\ref{S:2.2}.
%

%% 5. results and discussions
\section{Results and discussions}
\label{S:5}
\subsection{Apparent friction angle and impact area}
\label{S:5.1}
% figure-scalar sobol results
\begin{figure}[tpb]
\centering
\includegraphics[width=\textwidth]{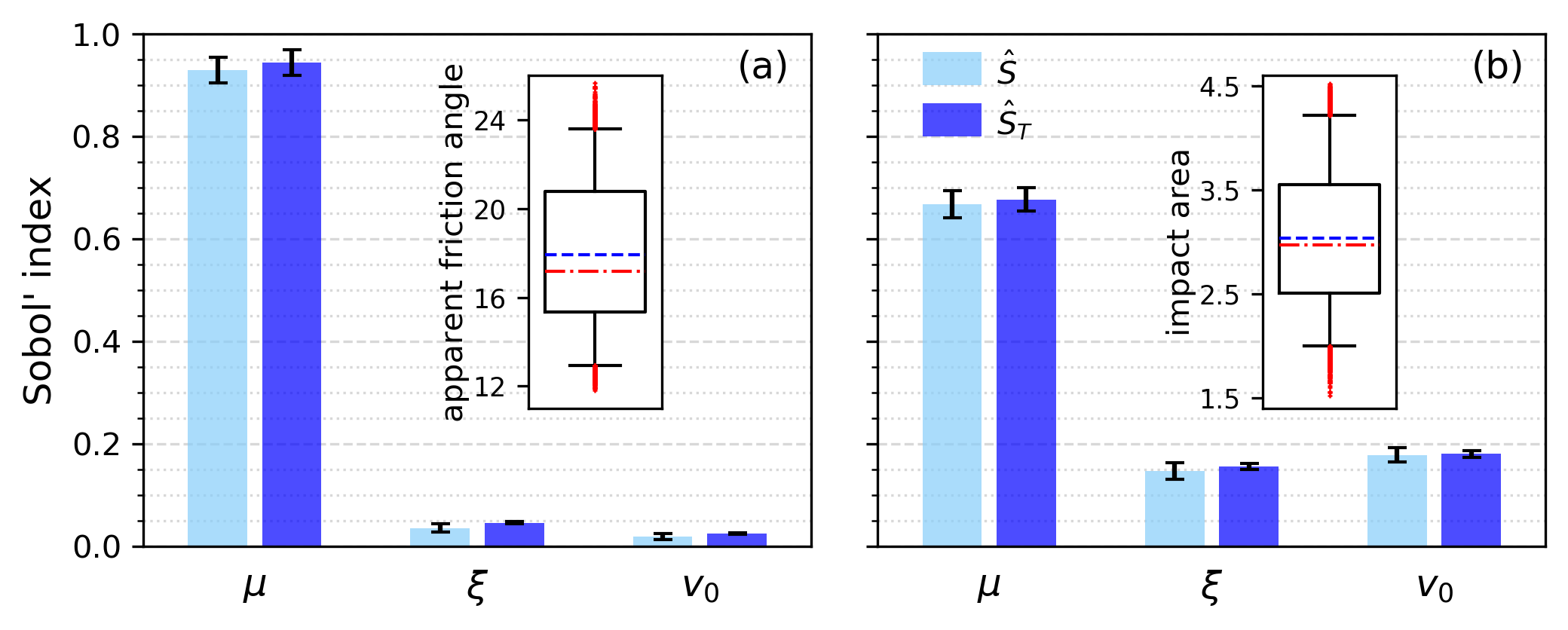}
\caption{Sobol' indices for aggregated scalar outputs (a) apparent friction angle and (b) impact area. The error bars of the bar plots indicate 95\% credible intervals of estimated Sobol' indices, which account for Monte Carlo-based sampling uncertainty and emulator uncertainty. The box plots show the distribution of emulator-predicted apparent friction angle values (in degree) and that of emulator-predicted impact area values (in Mio $m^2$). They represent the variation of the apparent friction angle and impact area resulting from the uncertain input variables respectively.}
\label{fig:sobol-scalar}
\end{figure}
The box plot in figure~\ref{fig:sobol-scalar}(a) shows the distribution of emulator-predicted apparent friction angle values corresponding to the 12000 samples used to estimate the variance of the apparent friction angle (see section~\ref{S:4.4}). Due to input uncertainties, the apparent friction angle could vary in a wide range, around $11.8^\circ$-$25.7^\circ$. The mean is $17.9^\circ$. The standard deviation is $3.1^\circ$ which corresponds to the square root value of $V(y)$ in eqs.~(\ref{eq:Si})-(\ref{eq:STi}). The bar plots in figure~\ref{fig:sobol-scalar}(a) display the estimated first-order and total-effect Sobol' indices, with CI(95\%) denoting the Monte Carlo-based sampling uncertainty and emulator uncertainty. Each pair of bar plots corresponds to the first-order and total-effect Sobol' indices of one input variable. It is evident that the apparent friction angle is dominated by the dry-Coulomb friction coefficient $\mu$ of which the first-order index is over 0.9, whereas both the turbulent friction coefficient $\xi$ and the release volume $v_0$ show little influence on the apparent friction angle, with both first-order indices being smaller than 0.05. This result is expected since $\mu$ governs the slope angle on which flow mass begins to deposit \citep{McDougall2017}, and it is consistent with the results based on one-at-a-time sensitivity analysis methods \citep[e.g.,][]{Schraml2015, Frey2016}. Furthermore, it is noteworthy that the difference between the first-order and total-effect indices is small, indicating weak interactions among the three input variables regarding the apparent friction angle.

Similarly, the box plot in figure~\ref{fig:sobol-scalar}(b) shows the distribution of emulator-predicted impact area values. Owing to input uncertainties, the impact area could vary between 1.5-4.5 Mio $m^2$ with a standard deviation 0.6 Mio $m^2$. From the bar plots, it can be seen that estimated first-order indices of $\mu$, $\xi$, and $v_0$ are around 0.67, 0.15, 0.18 respectively. It indicates that $\mu$ contributes the most to the variance of the impact area, followed by $v_0$ and $\xi$. Similar to the results on the apparent friction angle, the small difference between the first-order and total-effect indices implies that the three input variables barely interact with each other concerning the impact area. Compared to the results of the apparent friction angle, the importance of $\mu$ on the impact area decreases and that of $\xi$ and $v_0$ increases. A plausible explanation is that the apparent friction angle only depends on the deposit (assuming that the release area remains the same) where $\mu$ plays the dominant role, whereas the impact area depends on all inundated region where all three input variables may have impact.

\subsection{Maximum flow height and velocity}
\label{S:5.2}
% figure-statistics maxh and maxv
\begin{figure}[tpb]
\centering
\includegraphics[width=\textwidth]{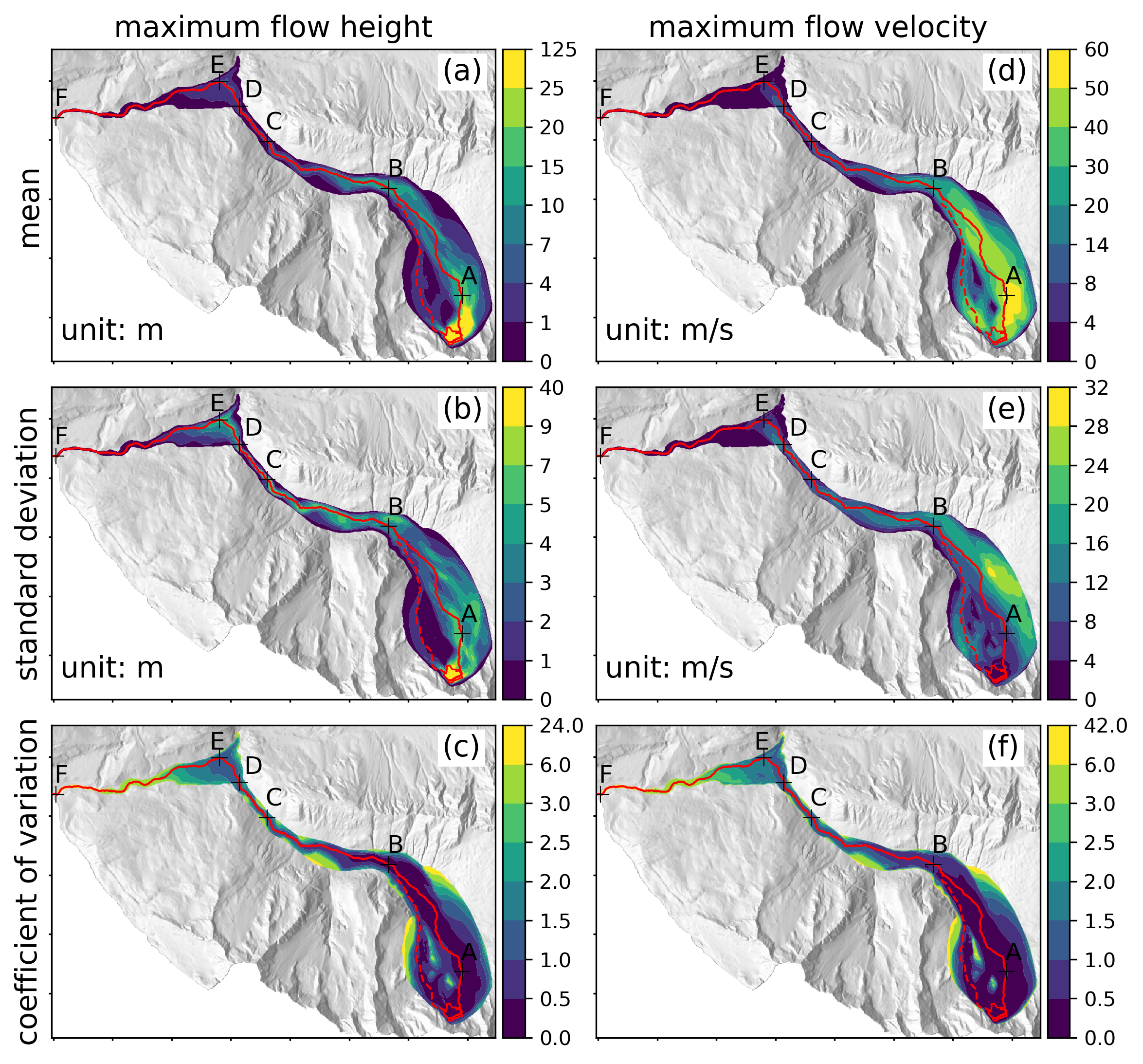}
%\makebox[\textwidth][c]{\includegraphics[width=1.05\textwidth]{figs//statistics_maxh_maxv.png}}
\caption{Statistics of emulator-predicted maximum flow height (left column) and velocity (right column) at $k=47958$ locations. For each location, the mean (first row), standard deviation (second row), and coefficient of variation (third row) are calculated from 12000 emulator-predicted maximum flow height and velocity values at that location (see section~\ref{S:4.4}). The polygon at the bottom-right corner of each panel denotes the release area. The local low/high values on the left side of location A in each panel result from the local ridges (see figure~\ref{fig:cengalo}). }
\label{fig:statistics-maxhv}
\end{figure}
Before discussing global sensitivity analysis results on maximum flow height and velocity, we summarize the statistics that are needed to interpret the results. Figures~\ref{fig:statistics-maxhv}(a)-(c) show the mean, standard deviation, and coefficient of variation of emulator-predicted maximum flow height values at each location. Figures~\ref{fig:statistics-maxhv}(d)-(f) show the counterparts of emulator-predicted maximum flow velocity values. The major and minor flow paths as well as locations A-F along the major flow path are noted to facilitate the description of results. The profile of the major flow path and the apparent friction angle values corresponding to locations A-F are shown in figure~\ref{fig:cengalo}. Location A sits near the release area, where the slope is steep. From location B to location D is the Bondasca valley. Location C corresponds to the mean location of 12000 apparent friction angle values ($17.9^\circ$), denoting the average run-out distance. From location D to location E is the debris flow retention basin \citep{Walter2020}. Location F is near the west boundary of the DEM. 

It can be seen from figures~\ref{fig:statistics-maxhv}(a) and (d) that in general, the mean of maximum flow height gradually decreases along the flow path whereas the mean of maximum flow velocity first increases then decreases reflecting the acceleration and deceleration process. Along the path cross section direction, both the mean of maximum flow height and that of maximum flow velocity generally decrease from the center to the sides. In addition, the mean values in the upstream area of location B are on average much larger than the mean values in the downstream area of location B, possibly because the average slope from the release zone to location B is larger than that beyond location B (see figure~\ref{fig:cengalo}) and the corner around location B decelerates the flow mass. 

The standard deviation shown in figures~\ref{fig:statistics-maxhv}(b) and (e) reflects the variation of maximum flow height and velocity at each location resulting from uncertainties of the three input variables. It corresponds to the square root of $V(y)$ in eqs.~(\ref{eq:Si})-(\ref{eq:STi}). In the Bondasca valley between location B and location D, where the channel is well-defined, the standard deviation generally decreases from the center to the sides in lateral direction, similar to the trend observed in figures~\ref{fig:statistics-maxhv}(a) and (d).

Figures~\ref{fig:statistics-maxhv}(c) and (f) present the coefficient of variation defined as the ratio of the standard deviation to the mean, representing the relative variation. Comparing figures~\ref{fig:statistics-maxhv}(c) and (f) with figures~\ref{fig:statistics-maxhv}(a) and (d), we find strong negative correlation between the coefficient of variation and the mean. The coefficient of variation generally increases both along the longitudinal direction and from the center to the sides in the lateral direction. A noteworthy feature is that figure~\ref{fig:statistics-maxhv}(b) shows large differences to figure~\ref{fig:statistics-maxhv}(e), whereas figures~\ref{fig:statistics-maxhv}(c) and (f) greatly resemble each other. It indicates that for maximum flow height and velocity, their absolute variation represented by the standard deviation differs from each other, whereas their relative variation represented by the coefficient of variation shows great similarities.

%figure-first order sobol indices for maxh and maxv
\begin{figure}[tpb]
\centering
\includegraphics[width=\textwidth]{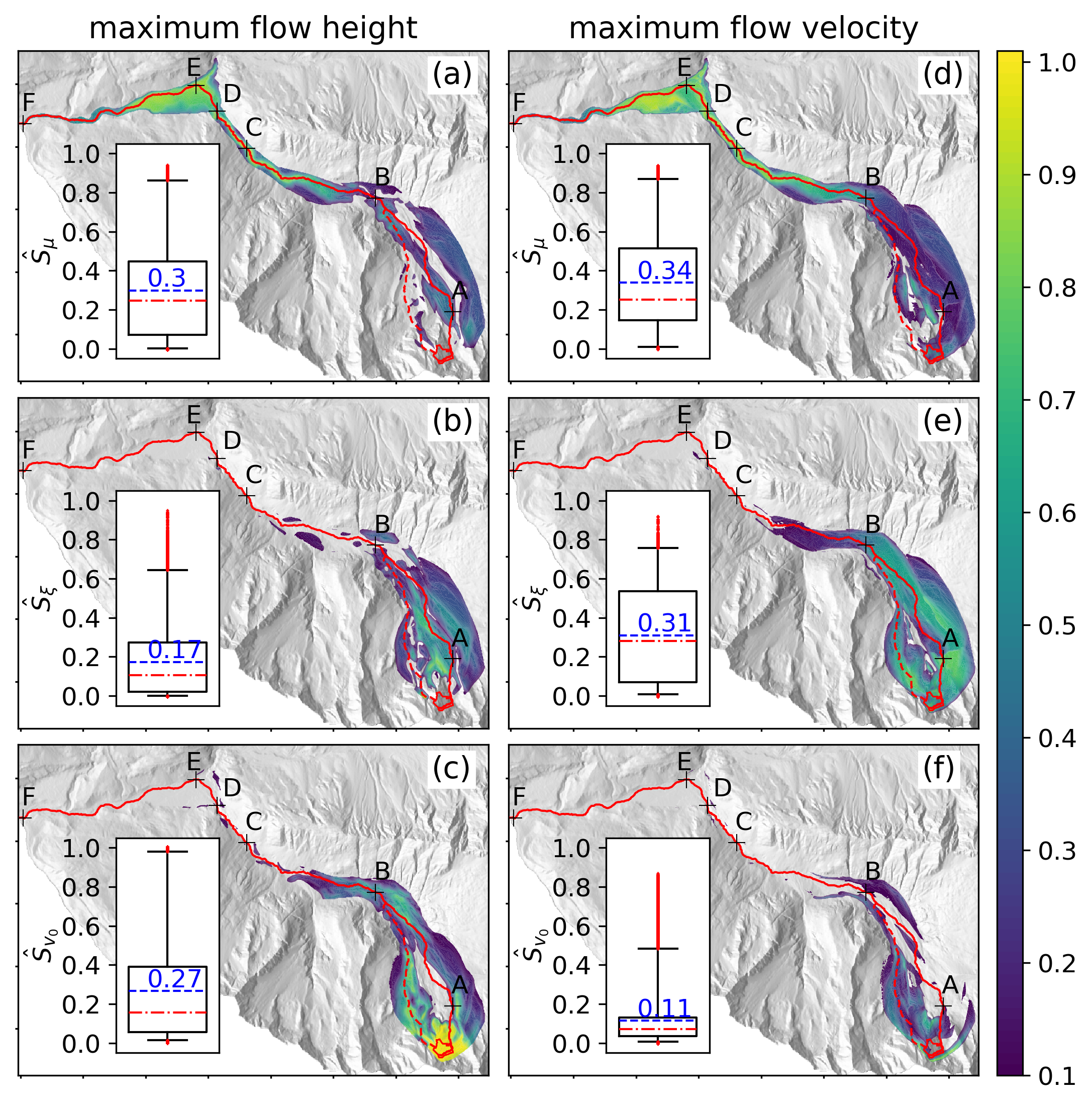}
%\makebox[\textwidth][c]{\includegraphics[width=1.05\textwidth]{figs//S_maxh_maxv.png}}
\caption{First-order Sobol' indices for $(h_{l_1}^{\text{max}}$,\ldots,$h_{l_k}^{\text{max}})^T$ (left column) and for $(\|\mathbf{u}_{l_1}\|^{\text{max}}$,\ldots,$\|\mathbf{u}_{l_k}\|^{\text{max}})^T$ (right column). In each panel, values smaller than 0.1 are not shown in the colormap; the box plot presents the distribution of respective first-order indices at all locations (including values smaller than 0.1); the mean over all locations is notated in the box plot.}
\label{fig:first-order-maxhv}
\end{figure}

\begin{figure}[tpb]
\centering
\includegraphics[width=\textwidth]{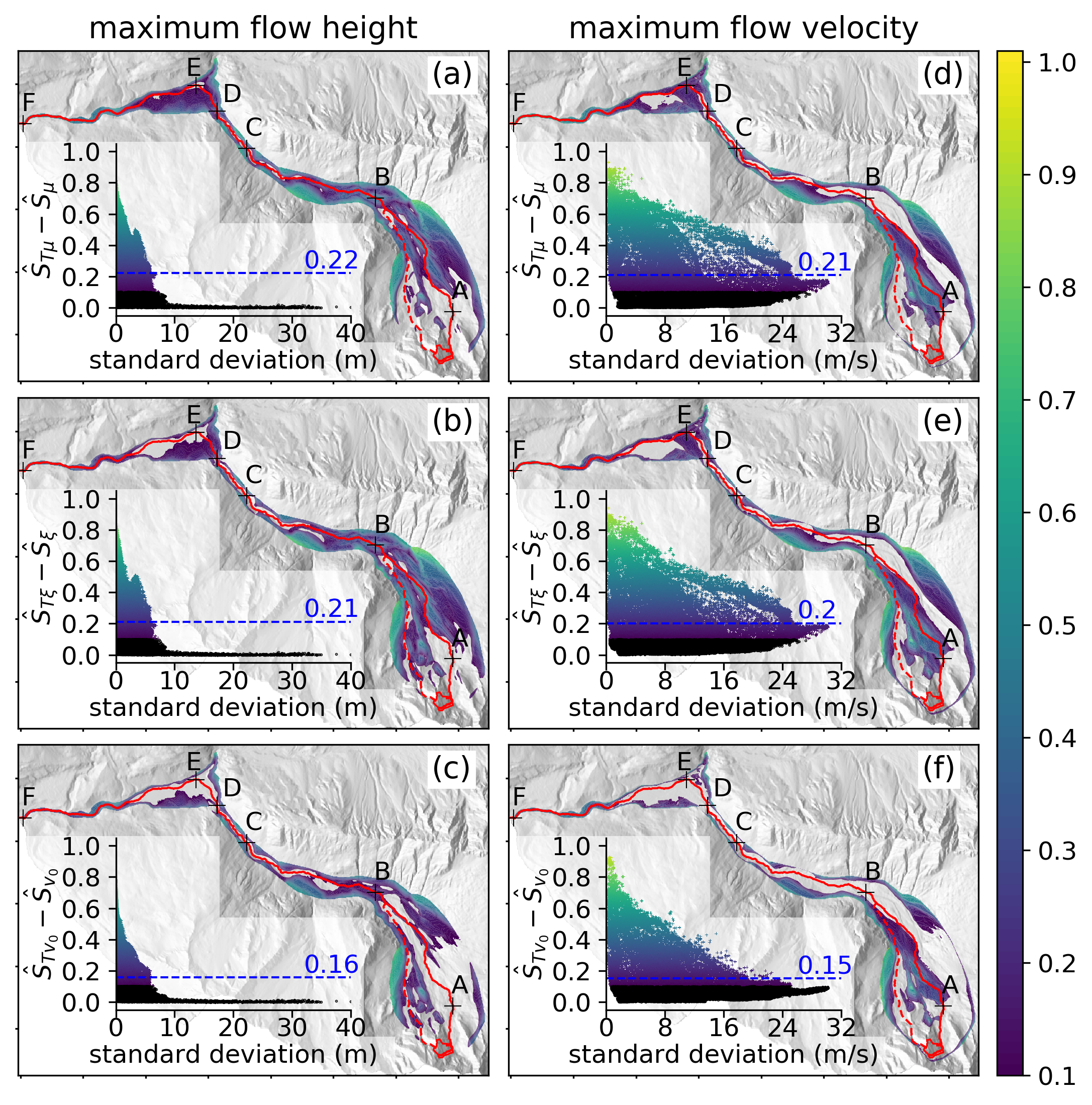}
%\makebox[\textwidth][c]{\includegraphics[width=1.05\textwidth]{figs//ST_minus_S_maxh_maxv.png}}
\caption{Difference between total-effect and first-order Sobol' indices for $(h_{l_1}^{\text{max}}$,\ldots,$h_{l_k}^{\text{max}})^T$ (left column) and for $(\|\mathbf{u}_{l_1}\|^{\text{max}}$,\ldots,$\|\mathbf{u}_{l_k}\|^{\text{max}})^T$ (right column). In each panel, values smaller than 0.1 are not shown in the colormap; the scatter plot shows the difference versus the standard deviation shown in figure~\ref{fig:statistics-maxhv}(b) and (e), where difference values larger than 0.1 are plotted using the same colorbar as that used for the colormap, and difference values smaller than 0.1 are plotted in black; the mean over all locations is notated in the scatter plot.}
\label{fig:high-order-maxhv}
\end{figure}

Figures~\ref{fig:first-order-maxhv}-\ref{fig:high-order-maxhv} present results of the Sobol' sensitivity analysis on maximum flow height and velocity at each location. The uncertainties of estimated Sobol' indices are found to be negligible and have little impact on the discussion (see figure~\ref{fig:sobol-scalar}). The CI(95\%) is therefore omitted here to avoid redundance. In addition, values smaller than 0.1 are not shown in the colormaps to highlight the trends that we will shortly discuss.

Figures~\ref{fig:first-order-maxhv}(a)-(c) show the first-order contributions of $\mu$, $\xi$, and $v_0$ to the variation of maximum flow height at each location. The mean values of $\hat{S}_\mu$, $\hat{S}_\xi$, and $\hat{S}_{v_0}$ over the 47958 locations are 0.3, 0.17, and 0.27 respectively. A closer look shows that the dry-Coulomb friction coefficient $\mu$ dominates in the downstream area beyond location B, whereas its impact in the upstream area of location B is limited; the turbulent friction coefficient $\xi$ is an influential factor in the upstream area of location B especially in areas around the major flow path, whereas it has negligible impact in the downstream area of location B; the release volume $v_0$ contributes the most in areas surrounding the release zone and has significant impact in areas near the minor flow path as well as areas surrounding location B, whereas it shows little influence in the downstream area similar as $\xi$. 

Figures~\ref{fig:first-order-maxhv}(d)-(f) present the first-order contributions of $\mu$, $\xi$, and $v_0$ to the variation of maximum flow velocity at each location. The mean values of $\hat{S}_\mu$, $\hat{S}_\xi$, and $\hat{S}_{v_0}$ over all the locations are 0.34, 0.31, and 0.11 respectively. A closer inspection shows that the variation of maximum flow velocity in the downstream area beyond location B is predominantly driven by $\mu$, while it has mild impact in the upstream area; $\xi$ contributes the most to the variation of maximum flow velocity in the upstream area of location B, where the mean values of maximum flow velocity are large (comparing figure~\ref{fig:first-order-maxhv}(e) with figure~\ref{fig:statistics-maxhv}(d)); $v_0$ only has mild impact in areas near the release zone and near the minor flow path. 

Comparing figures~\ref{fig:first-order-maxhv}(a)-(c) with figures~\ref{fig:first-order-maxhv}(d)-(f), we find the first-order contribution of $\mu$ to the variation of maximum flow height only slightly differs from its contribution to the variation of maximum flow velocity, with the mean over all locations increasing from 0.3 to 0.34; $\xi$ has more impact on maximum flow velocity than on maximum flow height, with a difference 0.14 on average; the influence of $v_0$ on maximum flow height is more important than its influence on maximum flow velocity, with a difference 0.16 on average. The dominant role of $\mu$ in the downstream area agrees with the finding in section \ref{S:5.1} that $\mu$ predominantly affects the apparent friction angle. The importance of $\xi$ in the upstream area with large mean values of maximum flow velocity is in accord with expectation since the turbulent friction term in eq.~(\ref{eq:vsmodel}) is proportional to the square of flow velocity.

% figure-high order sobol indices for maxh and maxv

Figures~\ref{fig:high-order-maxhv}(a)-(c) show the difference between total-effect and first-order Sobol' indices for maximum flow height at each location, which indicates the interactions between different input variables. Taking $\hat{S}_{T\mu}-\hat{S}_\mu$ as an example, it accounts for all high-order effects related to $\mu$, including the second-order interaction between $\mu$ and $\xi$, the second-order interaction between $\mu$ and $v_0$, as well as the third-order interaction among $\mu$, $\xi$, and $v_0$. The mean values of $\hat{S}_{T\mu}-\hat{S}_\mu$, $\hat{S}_{T\xi}-\hat{S}_\xi$, and $\hat{S}_{Tv_0}-\hat{S}_{v_0}$ over all locations are 0.22, 0.21, and 0.16 respectively. The areas showing significant difference coincide with the areas with low mean values, low standard deviation values, and high coefficient of variation values (see figure~\ref{fig:statistics-maxhv}(a)-(c)), except the area around the major flow path between location A and location B. The difference between $\hat{S}_{Tv_0}$ and $\hat{S}_{v_0}$ in this area is negligible, meaning that all high-order effects related to $v_0$ in this area are negligible. The difference in this area shown in figures~\ref{fig:high-order-maxhv}(a) and (b) is therefore mainly due to the interaction between $\mu$ and $\xi$. From the scatter plots of respective difference versus the standard deviation, it is evident that the interactions generally decrease with increasing standard deviation.

Figures~\ref{fig:high-order-maxhv}(d)-(f) show the difference between total-effect and first-order Sobol' indices for maximum flow velocity at each location. The mean values of $\hat{S}_{T\mu}-\hat{S}_\mu$, $\hat{S}_{T\xi}-\hat{S}_\xi$, and $\hat{S}_{Tv_0}-\hat{S}_{v_0}$ over all locations are 0.21, 0.2, and 0.15 respectively. Similar to the results on maximum flow height, the areas showing significant difference greatly resemble the areas with low mean values, low standard deviation values, and high coefficient of variation values of maximum flow velocity, see figures~\ref{fig:statistics-maxhv}(d)-(f). Again the area around the major flow path between location A and location B is an exception. It can be clearly seen from the scatter plots of respective difference versus the standard deviation, that the interactions generally decrease with increasing standard deviation.

Comparing figures~\ref{fig:high-order-maxhv}(a)-(c) with figures~\ref{fig:high-order-maxhv}(d)-(f), we find that for both maximum flow height and maximum flow velocity, most of the significant interactions occur on the margins of the flow paths where mean values and standard deviation values are relatively small, whereas values of coefficient of variation are relatively large (see figure~\ref{fig:statistics-maxhv}); the interactions generally decrease with increasing standard deviation; there are stronger interactions between the two friction coefficients $\mu$ and $\xi$ than between the release volume $v_0$ and each friction coefficient.

%% 6. Conclusions
\section{Conclusions}
\label{S:6}
In this study, we have presented a computationally efficient approach which enables variance-based global sensitivity analyses of computationally demanding landslide run-out models. The methodology couples the novel open-source mass flow simulation tool r.avaflow \citep{Mergili2017}, robust Gaussian process emulation for multi-output models \citep{Gu2016,Gu2018,Gu2019}, and a recent algorithm addressing the emulator uncertainty \citep{LeGratiet2014}. Based on the 2017 Bondo landslide event, we have employed the approach to study the global sensitivity of selected run-out model outputs to three input variables, namely the release volume and the two friction coefficients. Our main findings are as follows.  

\begin{enumerate}[$\bullet$]
\item The proposed approach can be successfully used to study the relative importance and interactions of input variables in landslide run-out models, when the trained Gaussian process emulators are validated and the base sample size of a Sobol' analysis is properly chosen.
\item The first-order effects of each input variable are broadly in line with results of common one-at-a-time sensitivity analyses in the literature. The dry-Coulomb friction coefficient dominates the apparent friction angle, as well as maximum flow height and velocity in the downstream area. The turbulent friction coefficient contributes the most to the variation of maximum flow velocity in the area where maximum flow velocity values are expected to be large. The release volume is found to have significant impact on maximum flow height in the area surrounding the release zone whereas it shows little impact on maximum flow velocity.
\item Interactions between the input variables could be analyzed for the full flow path, which cannot be assessed by commonly used one-at-a-time approaches. Significant interactions between the input variables generally happen on the margins of the flow path. The mean values and standard deviation values of maximum flow height and velocity are small in those areas. The interactions generally decrease with increasing variation of maximum flow height and velocity. Furthermore, there are stronger interactions between the two friction coefficients than between the release volume and each friction coefficient. 
\end{enumerate}

The proposed methodology can be easily extended for variance-based global sensitivity analysis on landslide run-out models employing other basal rheologies, or potentially on any computationally demanding models, when the assumption of Gaussian process emulation is fulfilled as stated in section \ref{S:2.3}. 

In addition, other computationally expensive tasks can also benefit from the significant speed-up owing to emulation techniques. While the run-out simulation takes 32 minutes on average to determine maximum flow height at the 47958 locations for a given parameter setting, this time reduces to 0.05 seconds for evaluating the emulator. Hence, whenever an application requires a large number of model evaluations, like uncertainty quantification and model calibration of landslide run-out models, computational costs for training the emulator will be compensated. In our study, this threshold is determined by the 200 training simulation runs, around 107 hours. The emulation techniques likewise have a great potential whenever a splitting between off-line computation (e.g. emulator training) and on-line computation (e.g. urgent computing for early warning systems) is feasible.

\section*{Acknowledgement}
The authors gratefully acknowledge the support of Hu Zhao by the China Scholarship Council (grant number: 201706260262) and by the Helmholtz Graduate School for Data Science in Life, Earth and Energy.

%\appendix
%\clearpage
%\input{appendix}


%% References


%\clearpage
%\bibliographystyle{model1-num-names.bst}
\bibliographystyle{model2-names.bst}\biboptions{authoryear}
\bibliography{global-sensitivity}






% \end{thebibliography}

\end{document}